\documentclass[preprint,5p,twocolumn]{elsarticle}

\usepackage[utf8]{inputenc}
\usepackage{slashed}
\usepackage{amssymb}
\usepackage{amsmath}
\usepackage{mathrsfs}
\usepackage{natbib}
\usepackage{soul}
\biboptions{sort&compress}
\usepackage{hyperref}

\usepackage{epsfig}
\usepackage{graphics}
\usepackage{float}
\usepackage{tabularx}
\usepackage{color}

\def\be{\begin{equation}}
\def\ee{\end{equation}}

\def\ba{\begin{eqnarray}}
\def\ea{\end{eqnarray}}
\def\beas{\begin{eqnarray*}}
\def\eeas{\end{eqnarray*}}

\begin{document}

\title{Plateau Inflation in $R$-parity Violating MSSM}

\author[PRL]{Girish Kumar Chakravarty}
\ead{girish20@prl.res.in}

\author[PRL]{Ujjal Kumar Dey}
\ead{ujjaldey@prl.res.in}

\author[DFCU,INFN]{Gaetano Lambiase}
\ead{lambiase@sa.infn.it }

\author[PRL]{Subhendra Mohanty}
\ead{mohanty@prl.res.in}

\address[PRL]{Physical Research Laboratory, Ahmedabad-380009, India}
\address[DFCU]{Dipartimento di Fisica ``E.R. Caianiello" Universit\'a di Salerno, I-84084 Fisciano (Sa), Italy}
\address[INFN]{INFN - Gruppo Collegato di Salerno, Italy}

\begin{abstract} 
Inflation with plateau potentials give the best fit to the CMB observables as they predict tensor to scalar ratio stringently bounded by the observations from Planck and BICEP2/Keck. In supergravity models  it is possible to obtain plateau potentials for scalar fields in the Einstein frame which can serve as the inflation potential by considering higher dimensional Planck suppressed operators and  by the choice of non-canonical K\"ahler potentials. We construct a plateau inflation model in MSSM where the inflation occurs along a sneutrino-Higgs flat direction. A hidden sector Polonyi field is used for the breaking of supersymmetry after the end of the inflation. The proper choice of superpotential leads to strong stabilization of the Polonyi field, $m_{Z}^2\gg m_{3/2}^2$, which is required to solve the cosmological moduli problem. Also, the SUSY breaking results in a TeV scale gravitino mass and scalar masses and gives rise to bilinear and triliear couplings of scalars which can be tested at the LHC. The sneutrino inflation field can be observed at the LHC as a TeV scale diphoton resonance like the one reported by CMS and ATLAS.  

\end{abstract}

\begin{keyword}
Inflation \sep Supersymmetry \sep Supergravity
\end{keyword}

\maketitle 

\section{Introduction}
There are atleast two experimental sectors which hint at the existence of scalar fields beyond the Higgs field of the Standard Model (SM). In order to explain the observed anisotropy of the cosmic microwave background (CMB) temperature at the super horizon scales \cite{Ade:2015xua} and at the same time the low value of the tensor-to-scalar ratio $r <0.07$ \cite{Ade:2015xua, Array:2015xqh} one requires an inflaton field  with a plateau potential \cite{Starobinsky:1979ty, Starobinsky:1980te, Ellis:2013xoa, Ellis:2014gxa, Kallosh:2013yoa, Kallosh:2015lwa, Chakravarty:2016fin}. The other hint for a scalar field is the 750 GeV diphoton excess which may have been observed by ATLAS \cite{Aaboud:2016tru} and CMS \cite{Khachatryan:2016hje} collaborations which has launched a large number of models which explain the 750 GeV diphoton resonance in the context of left-right models \cite{Dey:2015bur, Dasgupta:2015pbr, Dev:2015vjd, Deppisch:2016scs, Borah:2016uoi, Hati:2016thk, Agarwalla:2016rmw, Ghosh:2016lnu}, Grand Unification \cite{Patel:2015ulo, Aydemir:2016qqj, Nilles:2016bjl} and supersymmetry (SUSY) \cite{Chakraborty:2015gyj, Ding:2015rxx, Allanach:2015ixl, Choudhury:2016jbc, Djouadi:2016oey, Dreiner:2016wwk} and other exotic models (reviewed in \cite{Staub:2016dxq, Strumia:2016wys}). Cosmological implications of the 750 GeV scalar have been studied in \cite{Dhuria:2015ufo, Marzola:2015xbh, Ge:2016xcq, McDonald:2016cdh, Dimopoulos:2016tzn}. Sadly, the signal no longer persists as shown by the updated analysis of $\sqrt{s} = 13$ TeV data of ATLAS and CMS~\cite{Lenzi:2207286, CMS-PAS-EXO-16-027}.
In this paper we construct a plateau inflation model in the context of the minimal supersymmetric standard model which can be tested at the LHC in particular where the inflaton can be observed as a TeV scale diphoton resonance. We find that the most economical model which can explain both the phenomenon is to identify the left-handed sneutrino in a $R$-parity violating MSSM as the inflaton and as the diphoton resonance. The identification of the tau-sneutrino as the diphoton resonance has been made in the $R$-parity violating MSSM in \cite{Ding:2015rxx, Allanach:2015ixl}. On the other hand inflation with the singlet right-handed sneutrino has been well studied~\cite{Murayama:1992ua, Murayama:1993xu, Hamaguchi:2001gw, Ellis:2003sq, Antusch:2004hd, BasteroGil:2005gy, Kadota:2005mt} and in  MSSM the Higgs-sneutrino inflation along flat-directions \cite{Allahverdi:2005mz, Allahverdi:2006iq, Antusch:2010va, Antusch:2010mv, Pallis:2011ps, Haba:2011uz, Kim:2011ay, Khalil:2011kd, Aulakh:2012st, Arai:2012em, Nakayama:2013nya, Evans:2015mta, Deen:2016zfr} has also been studied. In this paper we consider a supergravity model with no-scale like K\"ahler potential and a superpotential which includes $R$-parity violating non-renormalizable operators at all orders. In this model we consider the supersymmetry breaking occurs via a Polonyi field which takes a non-zero vacuum expectation value after the end of inflation in the present epoch. The proper choice of superpotential in Polonyi field leads to much heavier Polonyi mass compared to gravitino mass to avoid the cosmological moduli problem and to obtain the vanishingly small cosmological constant $~10^{-120}$ \cite{Coughlan:1983ci, Ellis:1986zt, Goncharov:1984qm, Linde:2011ja, Dudas:2012wi}. The SUSY breaking generates masses which are of the TeV scale for all the SUSY scalar partners (like squarks, sneutinos and sleptons). The TeV scale sleptons are used in the loops for the production and decay of the TeV scale  sneutrino. The production and decay vertices which involve sneutrino-quark and sneutrino-sleptons are generated by the SUSY breaking by the hidden sector Polonyi field.

The rest of the paper is organized as follows. In Sec.~\ref{sec:model} we introduce the relevant K\"ahler potential and superpotential and construct the $D$-term and $F$-term potentials. In Sec.~\ref{sec:inflationDflat} we choose the $D$-flat Higgs-sneutrino direction that gives the required plateau potential from the $F$-term. We show how interaction terms arise from the Polonyi field SUSY breaking  which ultimately gives rise to the diphoton production and decay vertices in Sec.~\ref{sec:soft}. We then apply the model to the calculation of the $\sim$ TeV sneutrino production and decay and show the cross section for the diphoton resonance which can be a tentative signature for this model in Sec.~\ref{sec:diphoton}. We conclude and list future implications of the model in Sec.~\ref{sec:concl}.

\section{The model}
\label{sec:model}

In the early universe the no-scale K\"ahler potential gives the  plateau inflation from supergravity (SUGRA) \cite{Nakayama:2013txa, Ellis:2014gxa, Kallosh:2013yoa, Garg:2015mra, Chakravarty:2016fin} which fits the requirements of the observed low tensor-to-scalar ratio and the temperature anisotropy. We consider the most economical SUSY model, namely the MSSM but allow $R$-parity violation. In this model the inflaton is a linear combination of the sneutrino and the neutral components of the Higgs fields.

We consider $R$-parity violating terms in K\"ahler potential $K(\phi_{i},\phi^{*}_{i})$ 
\begin{align}
 K = 3 \ln \Big[1 + \frac{1}{3 M_{p}^{2}}\Big( &L^{\dagger}L + H^{\dagger}_{u} H_{u} + H^{\dagger}_{d}
 H_{d} + H^{\dagger}_{d}L \notag \\
  &+ L^{\dagger}
 H_{d} \Big)\Big] + Z Z^{\ast}-\frac{(Z Z^{\ast})^{2}}{\Lambda^2}  \label{KP1}
\end{align}
 and superpotential $W(\phi_{i})$ 
\begin{align}
W = \mu_{1} L \cdot &H_{u} + \mu_2 H_{u} \cdot H_{d} + \Delta M_{p}^{2} + \mu_{z}^{2} Z
 \nonumber \\
  &+ \frac{\lambda_{1}}{M_{p}} (L\cdot H_u)^2
 \exp\left(\frac{-L\cdot H_u}{M_{p}^{2}} \right) \nonumber \\ 
 & ~~ + \frac{\lambda_{2}}{M_{p}} (H_u\cdot H_d)^2
 \exp\left(\frac{H_u\cdot H_d}{M_{p}^{2}} \right).
 \label{SP1}
\end{align}
The noteworthy feature of the  superpotential (\ref{SP1}) is that it does not blow up even when the inflation fields are super-Planckian during inflation. The hidden sector Polonyi field $Z$ is introduced to break supersymmetry. The term $\left(-\frac{(Z Z^{\ast})^{2}}{\Lambda^2}\right)$ with $\Lambda \ll 1$ and the fine tuning of the constant $\Delta$ helps in the strong stabilization of the field $Z$ (i.e., $m_{Z}^2 \gg m_{3/2}^2$) and fixing the vanishingly small cosmological canstant $\sim 10^{-120}$ \cite{Linde:2011ja,Dudas:2012wi}. The other fields bear their standard meanings. In addition to the superpotential, given in Eq.~(\ref{SP1}), which we consider for the inflation we also consider the $R$-parity violating interaction terms
\be
W_{int}=\tilde{Y}_{ijk}L_{i}L_{j}e_{Rk} + \lambda^{\prime}_{ijk}L_{i}Q_{j}D_{k}
\label{SP2}
\ee
which will play a role in detection of the sneutrino at LHC.
From here onwards we shall work in the unit where $M_{p}=(8\pi G)^{-1}=1$.
The scalar potential in SUGRA depends upon the K\"ahler function $G(\phi_{i},\phi^{*}_{i})$ given in terms of superpotential 
$W(\phi_{i})$ and K\"ahler potential $K(\phi_{i},\phi^{*}_{i})$ as,
\be
G(\phi_{i},\phi^{*}_{i}) \equiv K(\phi_{i},\phi^{*}_{i}) + \ln W(\phi_{i}) +\ln W^{\ast}(\phi^{*}_{i}),
\ee
where $\phi_{i}$ are the chiral scalar superfields. 
The scalar potential is given as the sum of $F$-term and $D$-term potentials given by
\be
V_{F}=e^{G}\left[\frac{\partial G}{\partial \phi^{i}} K^{i}_{j*} \frac{\partial G}{\partial \phi^{*}_{j}} - 3 \right] \label{LV}
\ee
and
\be
V_D= \frac{1}{2}\left[\text{Re} f_{ab}^{-1}(\phi_{i})\right] D^{a}D^{b},\label{VD}
\ee
respectively, where $D^{a}=-g \frac{\partial G}{\partial \phi_{k}}(\tau^{a})_{k}^{l}\phi_{l}$ and $g$ is the gauge coupling constant corresponding to each gauge group and $\tau^{a}$ are corresponding generators. For $SU(2)_L$ symmetry $\tau^{a}=\sigma^{a}/2$, where $\sigma^{a}$ are Pauli matrices. For $U(1)_{Y}$ symmetry the hypercharges are $Y_u= 1$, $Y_d=-1$, $Y_L =-1$ for $H_u$, $H_d$, $L$ respectively. The quantity $f_{ab}$ is related to the kinetic energy of the gauge fields and is a holomorphic function of superfields $\phi_i$.

The kinetic term of the scalar fields is given by
\be
\mathcal{L}_{KE}=K_{i}^{j*} \partial_{\mu}\phi^{i} \partial^{\mu}\phi^{*}_{j}~, 
\label{LK}
\ee
where $K^{i}_{j*}$ is the inverse of the K\"ahler metric $K_{i}^{j*} \equiv \partial^{2}K / \partial\phi^{i}\partial\phi^{*}_{j}$.

Taking charged components of $SU(2)_L$ Higgs doublets $H_{u}$, $H_{d}$ to be zero in the classical background fields
during inflation. The $H_u$, $H_d$ and slepton doublet $L$
can be written as
\be
L =\begin{pmatrix}
  \phi_\nu\\
  0
 \end{pmatrix}\,,~~~~~~~~~
 H_{u} =\begin{pmatrix}
  0  \\
  \phi_{u}
 \end{pmatrix}\,,~~~~~~~~~
 H_{d} =\begin{pmatrix}
  \phi_{d}  \\
  0
 \end{pmatrix},\label{LHu}
\ee
and substituting in Eq.~(\ref{VD}) and assuming the canonical form of
the gauge kinetic function $f_{ab}=\delta_{ab}$, the $D$-term potential comes out to be
\begin{align}
V_D = &\frac{9}{8} \left(g_{1}^{2}+g_{2}^{2}\right) \nonumber\\ 
 & \times \frac{\left(-|\phi_u|^2+|\phi_d|^2+|\phi_\nu|^2+(\phi_d \phi_{\nu}^{\ast}
+\phi_{d}^{\ast} \phi_{\nu})/2\right)^{2}}{\left(3+|\phi_u|^2+|\phi_d|^2+|\phi_\nu|^2+(\phi_d \phi_{\nu}^{\ast}
+\phi_{d}^{\ast} \phi_{\nu})/2\right)^{2}} .
\end{align}
We choose the field configurations such that $V_D=0$ during inflation. Such a $D$-flat direction is given by the relation $\phi_\nu=\phi_{u}=\phi$, $\phi_d=0$. We next compute the $F$-term potential to study inflation.

\section{Inflation along $D$-flat direction}
\label{sec:inflationDflat}

With the assumption $\phi_\nu=\phi_{u}=\phi$, $\phi_d=0$, 
the K\"ahler potential (\ref{KP1}) and the superpotential (\ref{SP1}) reduce to the  simple forms 
\be
 K = 3 \ln \left[1+\frac{2\phi \phi^{*}}{3}\right]\,,\label{KP}
 \ee
\be
W= \mu_1 \phi^{2} + \lambda_1 \phi^{4} \exp(-\phi^2)\,.\label{SP}
\ee
During inflation we assume that SUSY is unbroken and the hidden sector field $Z=0$.
With the K\"ahler potential (\ref{KP}) and the superpotential (\ref{SP}), we get $F$-term scalar potential as
\begin{align}
V_F = \frac{\lambda_{1}^{2}}{243} & e^{-(\phi^{2}+\phi^{\ast 2})} \big(3+2 |\phi|^2\big)^3 |\phi|^6 \notag \\ 
& \times \left[84-7\left(\phi^{2}+\phi^{\ast 2}\big)\big(6+10|\phi|^2+4|\phi|^4\right) \right. \notag \\ 
 & ~~~~~~\left. + 149|\phi|^2+119|\phi|^4+26|\phi|^6+8|\phi|^8\right],\label{VFDflat}
\end{align}
where we have assumed that during inflation when the field values are of Planck scale, the linear term in superpotential 
(\ref{SP1}) does not contribute to
the inflation potential. Whereas in the later universe at SUSY breaking scale when the field values are at TeV scale, the 
higher order Planck suppressed terms become negligible compared to linear terms in $W$.


In order to obtain the inflationary observable predictions of the model, the kinetic term for the scalar field $\phi$ must be made
canonical. For K\"ahler potential (\ref{KP}) and superpotential (\ref{SP}), the kinetic term for $\phi$ field is obtained as
\be
\frac{18}{(3+2|\phi|^{2})^{2}}|\partial_{\mu}\phi|^{2},\label{Lke}
\ee
to make it canonical, we redefine the field $\phi$ to $\chi_{E}$ via
\be
\left|\partial_{\mu}\chi_{E} \right|^{2} = \frac{18}{(3+2\left|\phi\right|^{2})^{2}}\left|\partial_{\mu}\phi \right|^{2}, 
\label{chitophi1}
\ee
whose solution is given by
\be
\phi=\sqrt{\frac{3}{2}} \tan\left(\frac{\chi_{E}}{\sqrt{3}}\right)\,, \label{phisol}
\ee
decomposing $\phi$ in terms of its real and imaginary parts
$\phi=(\phi_{R}+i\phi_{I})/\sqrt{2}$ and assuming that during inflation the real part is
zero, we have $\phi=-\phi^{\ast}=i \phi_{I}/\sqrt{2}$.
Similarly, $\chi_{E}=(\chi_{R}+i\chi)/\sqrt{2}$ and fixing the real part to zero, we have
$\chi_{E}=-\chi_{E}^{\ast}=i \chi/\sqrt{2}$. 
The solution (\ref{phisol}) can be given as
\be
\phi_{I}=\sqrt{3} \tanh\left(\frac{\chi}{\sqrt{6}}\right)\,,\label{phichi}
\ee
where we have used the trigonometric relation $\tan(i \theta)=i \tanh(\theta)$.
Here $\chi=\sqrt{2}~{\rm Im}(\chi_E)$ plays the role of the inflaton and we kept the imaginary part~(\ref{phichi}) non-zero
as it can provide the required slow-roll potential. If we assume the real part of the field to be non-zero and imaginary
parts to be zero, we obtain a very steep potential due to $\phi\propto \tan(\chi_R)$ during inflationary 
regime and therefore the potential becomes unsuitable for slow-roll inflation.

Using Eq.~(\ref{phichi}), the scalar potential (\ref{VFDflat}) in terms of canonical 
field $\chi$ can be given as
\begin{align}
V_{F}(\chi) = &\frac{9}{32} \lambda_{1}^{2} \exp\left(3 \tanh\big(\chi/\sqrt{6}\big)^{2}\right) 
\tanh\big(\chi/\sqrt{6}\big)^{6} \nonumber \\
&\times \Big[1+\tanh\big(\chi/\sqrt{6}\big)^{2}\Big]^3 \Big[112 + 466 \tanh\big(\chi/\sqrt{6}\big)^{2} \notag \\
& ~~~~~~~ + 777 \tanh\big(\chi/\sqrt{6}\big)^{4} + 369 \tanh\big(\chi/\sqrt{6}\big)^{6} \notag \\
& ~~~~~~~ + 54 \tanh\big(\chi/\sqrt{6}\big)^{8} \Big]\,.
\label{Upot}
\end{align}

\begin{figure}[t!]
 \centering
\includegraphics[width= 8cm]{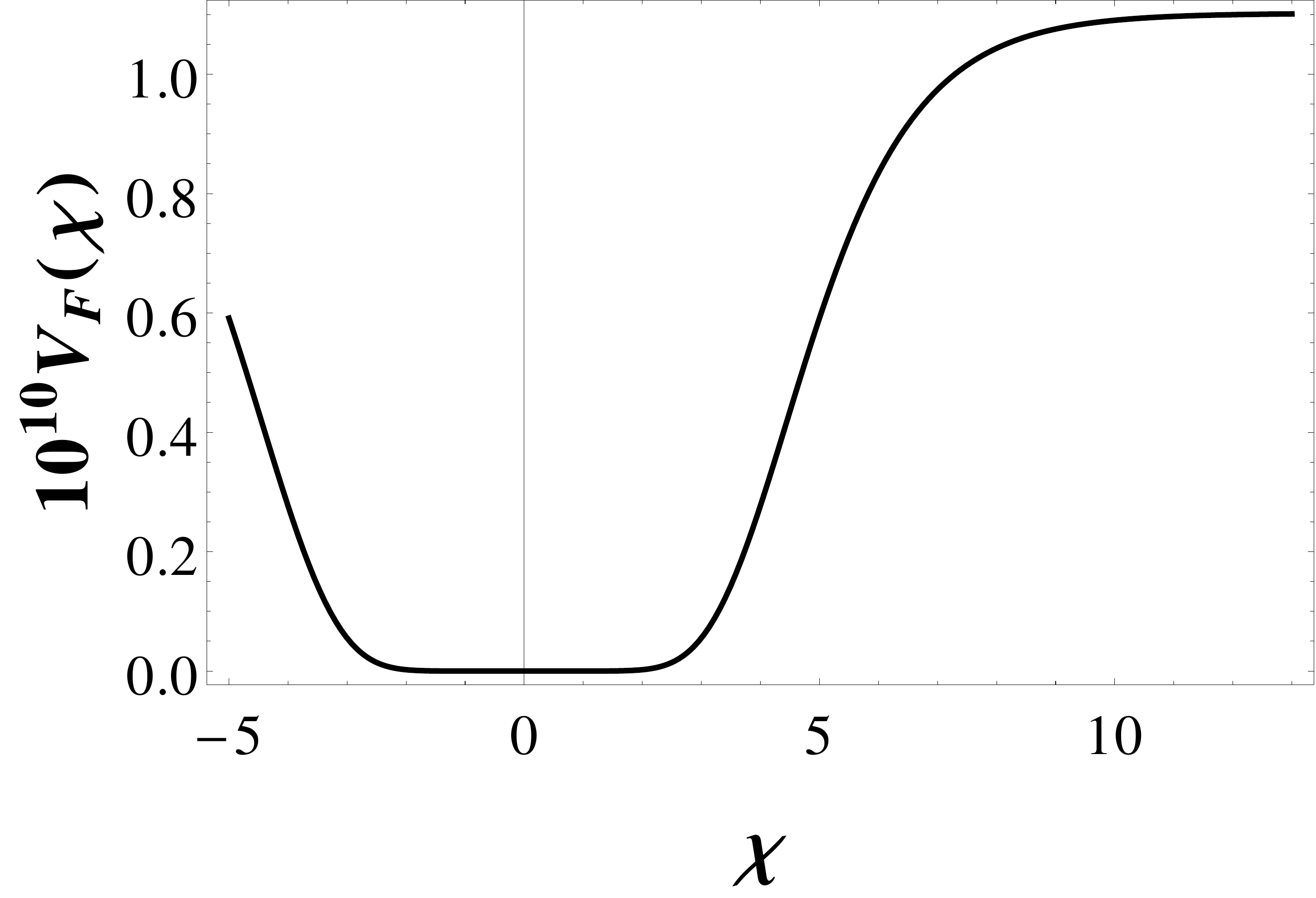}
\caption{The $D$-flat inflaton potential for which $\phi_\tau=\phi_d =0$ and $\phi_{\nu}=\phi_\nu=\phi$,
and evaluated at $\phi=\phi_I(\chi)$ and $Z\sim0$ is shown.}
 \label{fig2}
\end{figure}

Given the potential (\ref{Upot}), we can estimate the infaltionary observables using the
slow-roll parameters defined in the Einstein frame as
\be
\epsilon = \frac{1}{2}\left(\frac{V_{F}'}{V_{F}}\right)^2,
~~~~~~~\eta = \frac{V_{F}''}{V_{F}}\,,
~~~~~~~\xi = \frac{V_{F}'V_{F}'''}{V_{F}^{2}}.
\ee
We use the standard Einstein frame relations for the inflationary observables: amplitude of the curvature perturbation
$\Delta_{\mathcal R}^{2}$, tensor to scalar ratio $r$, 
scalar spectral index $n_{s}$ and running of spectral index $\alpha_s$, given by
\ba
\Delta_{\mathcal R}^{2} &=& \frac{1}{24 \pi^{2}} \frac{V_F}{\epsilon}\,,\\
r &=& 16\epsilon\,,\\
n_{s} &=& 1-6\epsilon+2\eta\,,\\
\alpha_{s} &\equiv& \frac{dn_{s}}{d\ln k} = 16\epsilon\eta -24\epsilon^{2} - 2\xi\,,
\ea
respectively.

For successful cosmology, it is required to have minimum $60$ $e$-folds of expansion during inflation when
the field value evolves from some initial value $\chi_s$ to some final value $\chi_e$. Field value $\chi_e$ at the end 
of inflation can be determined from the condition $\epsilon(\chi_e)=1$ for the end of inflation. The quantity $\chi_s$,
which corresponds to $N \approx 60$ $e$-folds before the end of inflation when observable CMB modes leave the horizon, can be
determined using the following $e$-folding expression
\be
N=\int_{\chi_e}^{\chi_s} \frac{V_F}{V_{F}'} d\chi\, .\label{efolds}
\ee 

The Planck-2015 analysis of CMB temperature and
polarization data combined with BICEP2/Keck Array CMB polarization observations have 
put an upper bound on tensor-to-scalar ratio $r_{0.05} < 0.07 \,(95\%$ CL)~\cite{Array:2015xqh}. 
Also the Planck observations give the scalar amplitude, the spectral index, the running of the spectral index as
$10^{10}\ln(\Delta_{\mathcal R}^{2}) = 3.089\pm 0.036$, $n_{s}= 0.9666 \pm 0.0062$,
$\alpha_s = - 0.0084 \pm 0.0082$, respectively, at ($68 \%$ CL, PlanckTT+lowP)~ \cite{Ade:2015xua, Ade:2015lrj}.

From the numerical analysis of the model with the theoretical and observational results for the CMB observables as discussed above,
we can fix the field values $\chi_s$, $\chi_e$ and the parameter $\lambda_1$ for $N\approx60$ $e$-folds.
We find that for $\chi_{s}\simeq8.96$ and $\lambda\simeq3.7\times10^{-8}$, we obtain $r \simeq 0.0033$, $n_{s} \simeq 0.9664$ and 
$\alpha_{s} \simeq -5.56\times10^{-4}$ consistent with the CMB observations. From the condition $\epsilon(\chi_e)=1$, we obtain
$\chi_e\simeq3.65$. For the rolling of the inflaton from $\chi_{s}\simeq8.96$ to $\chi_e\simeq3.65$, we obtain the minimum 
required $e$-folds $N\simeq60$. The inflaton potential along $D$-flat direction is shown in Fig.~\ref{fig2}.

\section{Soft SUSY breaking and scalar masses}
\label{sec:soft}
To study the inflationary dynamics, we assumed that the hidden sector field $Z=0$ and supersymmetry is unbroken at the time of inflation. After the end of inflation the scalar fields 
effectively become vanishing and $Z$ settles down to a finite minimum which results in SUSY breaking giving rise to the soft-breaking terms. To study the dynamics
of the $Z$ field and its minimization, we first make the kinetic term for the field canonical.
Assuming, at SUSY breaking, the scalar fields become
$\phi_{\nu}=\phi_\tau \sim0$ and $\phi_u =246\sin \beta$ GeV, $\phi_d =246 \cos \beta$ GeV, where $\beta = \tan^{-1}(\phi_{u}/ \phi_{d})$,
from (\ref{LK}), the dominant contribution to kinetic term for $Z$ comes out to be
\be
\left(1 -\frac{4}{\Lambda^2}|Z|^{2}\right)|\partial_{\mu}Z|^{2}\,
\ee
we find that for $\Lambda\ll 1$, field $Z$ behaves like a canonical field.

The $D$-term potential which is independent of the hidden sector field
$Z$ and vanishing for $\phi_{i}\sim0$, where $i=\nu, \tau, u, d$, does not contribute to the late time evolution of the universe, instead 
the effective potential is given by $F$-term. For deriving the $F$-term potential $V_F(\phi_i, Z)$, from Eq.~(\ref{LV}), in the present universe 
when $\phi_i \sim 0$ the dominant contribution comes from the linear terms in superpotential (\ref{SP1}) whereas the higher order terms in it 
with couplings $\lambda_1, \lambda_2$ can be neglected. There can be two possible scenarios, one with $\Delta=0$ and another with  $\Delta \neq 0$. In the first scenario, for the specific choice $(-1/\Lambda^{2}) = 2/\alpha$ where $\alpha= (1/4)+(3/16)\left[(2+\sqrt{3})^{1/3}+ (2-\sqrt{3})^{1/3}\right] \simeq 0.662$, Polonyi field $Z$ which breaks supersymmetry acquire a minima at $Z_{\rm min} \simeq 0.911$. The parameter $\alpha$ can be fine tuned near $\alpha\simeq0.662$ to obtain the cosmological constant $V(Z_{\rm min}) \lesssim 10^{-120}$. 
For $\mu_{S}\simeq1.6\times10^{-8}$, the gravitino mass is $m_{3/2}\sim 1~$TeV, and for $\mu_{1}, \mu_{2} \lesssim 10^{-17}$, the universal scalar masses are equal to gravitino mass $m_{\phi_i}^2 \simeq m_{3/2}^2$. However, this scenario is cosmologically not suitable because mass of the Polonyi field is of the same order as the gravitino mass $m_{Z}^2\sim m_{3/2}^2$. After inflation Polonyi field $Z$ does not settle to its minimum immediately instead it oscillates with large amplitude around its
potential minimum, the $\mathcal{O}({\rm TeV})$ scale oscillations of the field around the minima decays much after Big Bang Nucleosynthesis which leads to cosmological Polonyi problem \cite{Coughlan:1983ci,Ellis:1986zt,Goncharov:1984qm} (which is a special case of cosmological moduli problem \cite{Banks:1993en,deCarlos:1993wie}). However this problem can be solved if the Polonyi field is strongly stabilized i.e., $m_{Z}^2\gg m_{3/2}^2$ \cite{Linde:2011ja,Dudas:2012wi}.

The strong stabilization of the Polonyi field can be achieved with the second scenario $\Delta \neq 0$. For $\Lambda \ll 1$, we find the potential minimum $V_{\rm min}=- 3\Delta^2+\mu_{z}^{4}$ at $Z_{\rm min}=\frac{\Delta \Lambda^2}{2\mu_{z}^2}$. Therefore for $\mu_{z}^2 = \sqrt{3}\Delta$, the cosmological constant can be made as small as $\sim 10^{-120}$. The field at the minimum of potential becomes $Z_{\rm min}=\frac{\Lambda^2}{2\sqrt{3}}$ and the garvitino and $Z$ masses are obtained as 
\ba
m_{3/2}^{2}&=&\Delta^2\,,\\
m_{Z}^{2}&=&\frac{12\Delta^2}{\Lambda^2}=\frac{12 m_{3/2}^{2}}{\Lambda^2}\gg m_{3/2}^{2}\,,
\ea
respectively. For $\Lambda\sim10^{-2}$ and $\Delta\simeq 4\times 10^{-16}\sim 1$ TeV, we obtain $m_{3/2}\sim 1$ TeV and $m_{z}\sim\mathcal{O}(100~ {\rm TeV})$.

%
%

The scalar masses $m^{2}_{\phi_{i}} = \frac{\partial_{\phi}\partial_{\phi^{\ast}}V}{\partial_{\phi}\partial_{\phi^{\ast}}K}$ evaluated at $\phi_i\sim0$ and $Z_{\rm min}=\frac{\Lambda^2}{2\sqrt{3}}$, comes out to be
\begin{subequations}
\begin{align}
m^{2}_{\phi_{\nu}} &= \Delta^{2}+ \mu_{1}^{2}\,, \\
m^{2}_{\phi_{\tau}} &= \Delta^{2}\,,\\
m^{2}_{\phi_{u}} &= \Delta^{2}+ \frac{4}{3}(\mu_{1}^{2}+\mu_{1}\mu_{2}+\mu_{2}^2) \,, \\
m^{2}_{\phi_{d}} &= \Delta^{2}+ \mu_{2}^{2} \,.
\end{align}
\label{mphi}
\end{subequations}
Therefore, for $\mu_{1}^{2}, \mu_{2}^{2} \ll \Delta^{2}$, the 
scalar masses are equal to the gravitino mass $m_{3/2}\simeq m_{\phi_{i}}$.


Following Refs.~\cite{Soni:1983rm,Kaplunovsky:1993rd,Dudas:2005vv}, we calculate the coefficients of the 
soft SUSY breaking terms which arise from the K\"ahler potential (\ref{KP1}) and superpotential
(\ref{SP1}) and (\ref{SP2}). The effective potential of the observable scalar sector consists of soft mass terms which give 
(\ref{mphi}), and trilinear and bilinear soft SUSY breaking terms
\be
\frac{1}{3}A_{ijk}\phi^{i}\phi^{j}\phi^{k}
+\frac{1}{2}B_{ij}\phi^{i}\phi^{j}+ \rm{h.c.}\,,\nonumber
\ee
the coefficients $A_{ijk}$ and $B_{ij}$ are given by
\begin{subequations}
\begin{align}
\label{tri}
A_{ijk} &= F^{I}\partial_{I}Y_{ijk}+\frac{1}{2} F^{I} (\partial_{I} \hat K) Y_{ijk}  \,,\\
\label{bi} 
B_{ij} &= F^{I}\partial_{I}\mu_{ij}+\frac{1}{2} F^{I}(\partial_{I} \hat K) \mu_{ij} - m_{3/2} \mu_{ij}\,,
\end{align}
\end{subequations}
where the index $I$ is over the hidden sector fields and, the un-normalised masses $\mu_{ij}$
and Yukawa couplings $Y_{ijk}$ in terms of normalized ones are given by
\be
Y_{ijk}=e^{\hat{K}/2} \tilde Y_{ijk}\,;~~~~~~\mu_{ij}=e^{\hat{K}/2} \tilde \mu_{ij}\,,
\ee
and 
\be
F^{I}= e^{\hat{K}/2} \hat{K}^{IJ^{\ast}} \left(\partial_{J^{\ast}}\hat{W}^{\ast}
+\hat{W}^{\ast}\partial_{J^{\ast}}\hat{K}\right);~~
\hat{K}^{IJ^{\ast}}=(\hat{K}_{J^{\ast}I})^{-1}.
\ee
From Eqs.~(\ref{KP1}) and (\ref{SP1}), in this model, we have $\hat{K}(Z,Z^{\ast})=Z Z^{\ast}-
\frac{1}{\Lambda^2} (Z Z^{\ast})^{2}$ and $\hat{W}(Z)=\Delta + \mu_{z} Z$. The trilinear (\ref{tri}) and bilinear
(\ref{bi}) coefficients, up to $\Lambda\ll1$, are obtained as
\begin{align}
\label{eq:Aijk}
A_{ijk} &= \frac{\Lambda^2}{2\times3^{1/4}\sqrt{\Delta}} ~m_{3/2} \tilde Y_{ijk}\,,\\
\label{eq:Bij}
B_{ij} &= \left(\frac{\Lambda^2}{2\times3^{1/4}\sqrt{\Delta}}-1\right) m_{3/2} \tilde \mu_{ij}\,,
\end{align}
respectively.
We drop the $B_{ij}$ contribution to the scalar masses by taking corresponding $\tilde\mu_{ij}$ small. We will now study the phenomenological consequences of the universal scalar masses $m_0=m_{3/2} \sim {\rm TeV}$ and the scalar trilinear couplings 
(\ref{eq:Aijk}) at LHC.

\section{Observation at LHC}
\label{sec:diphoton}
\begin{figure}[!t]
 \centering
\includegraphics[scale=0.25]{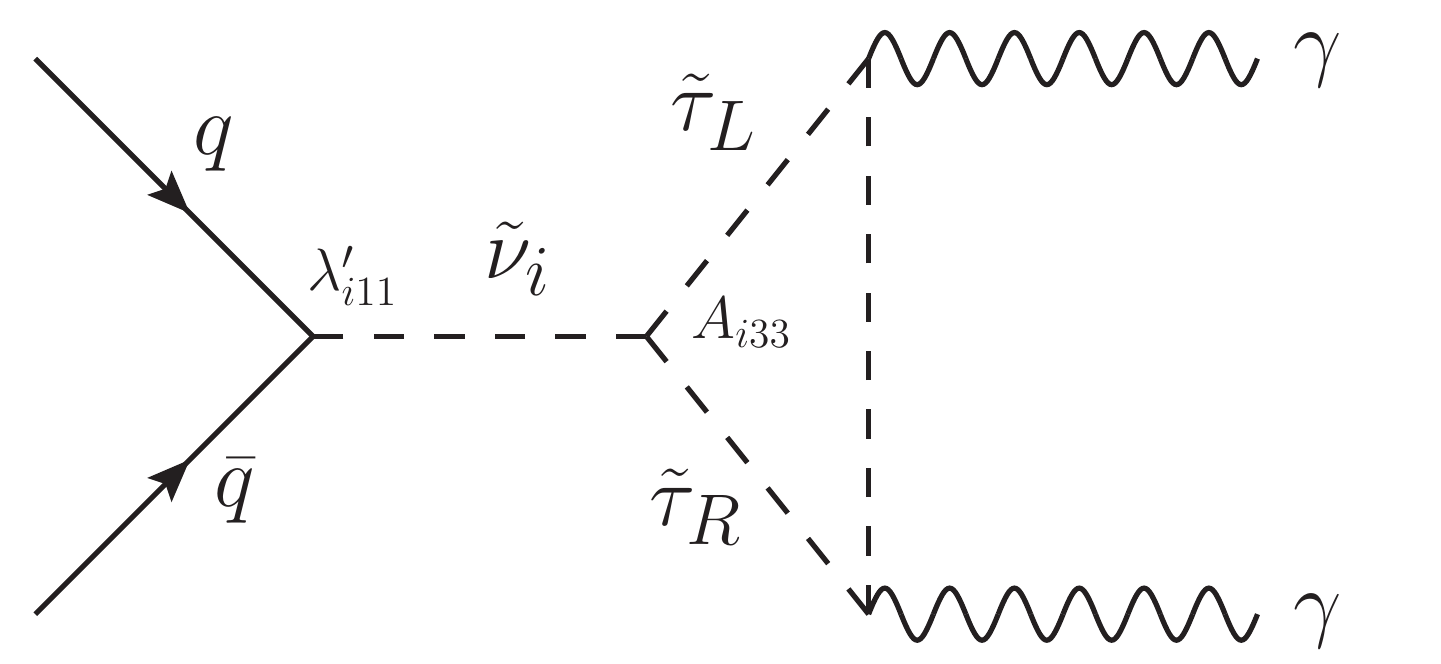}~~
\includegraphics[scale=0.25]{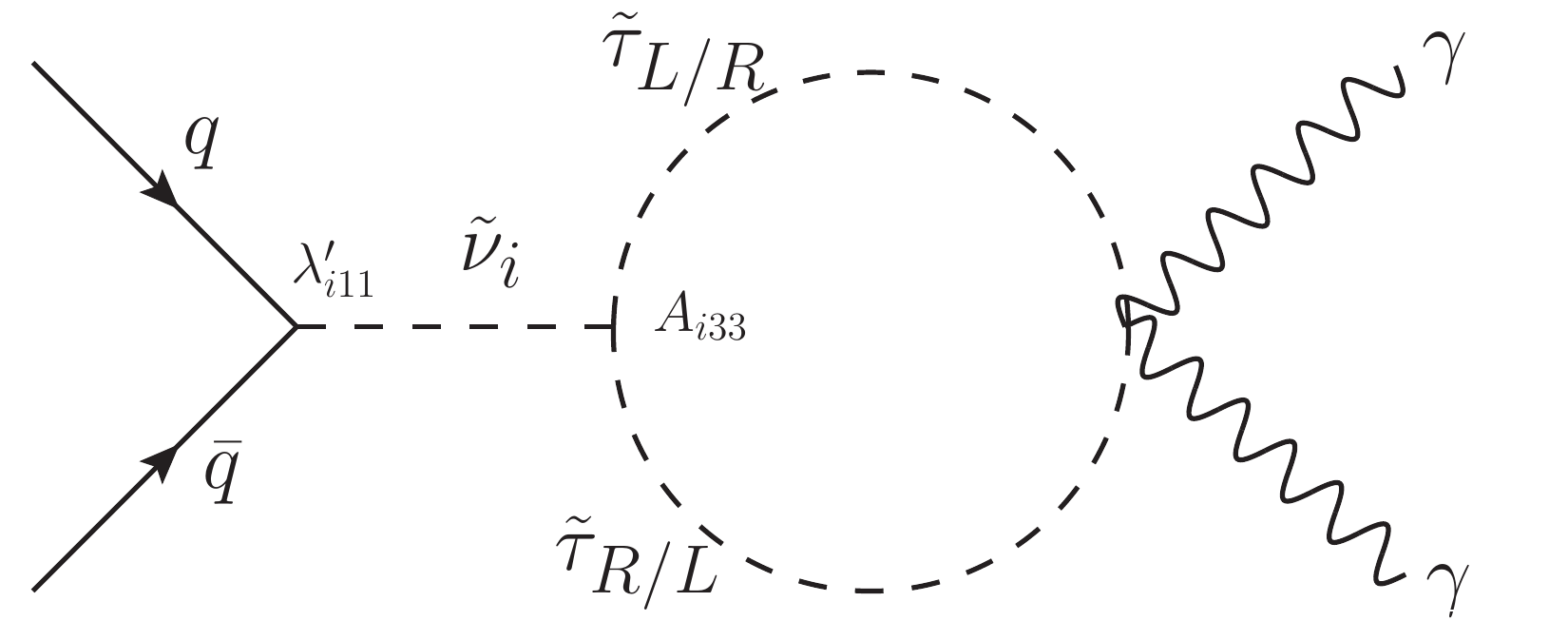}
\caption{Feynman diagrams for resonant sneutrino production and its subsequent decay to diphoton via $R$-parity
violating couplings.}
 \label{fig:feyndiag}
\end{figure}

\begin{figure}[!t]
 \centering
\includegraphics[scale=0.28]{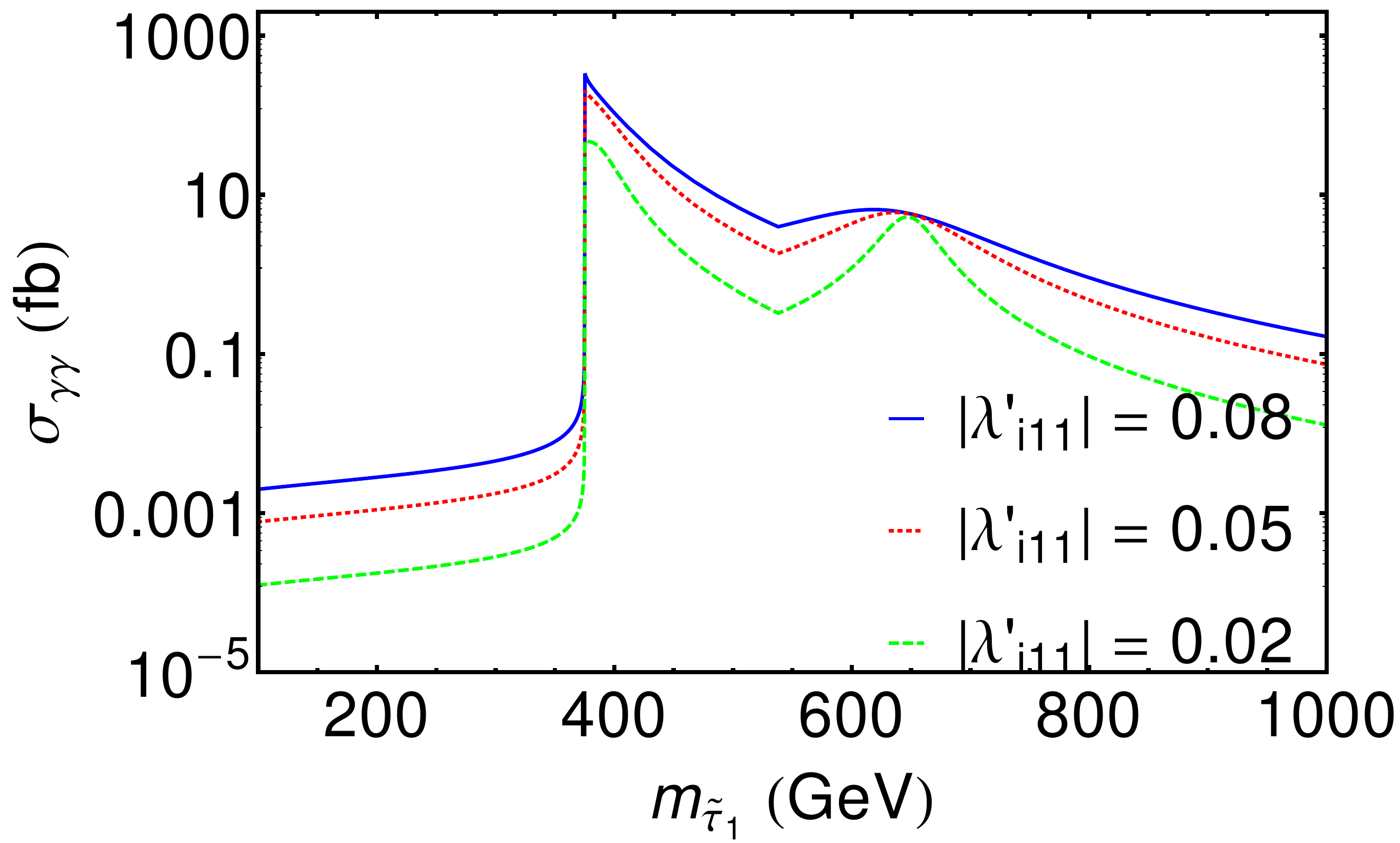} \\
\vspace*{0.5cm}
\includegraphics[scale=0.23]{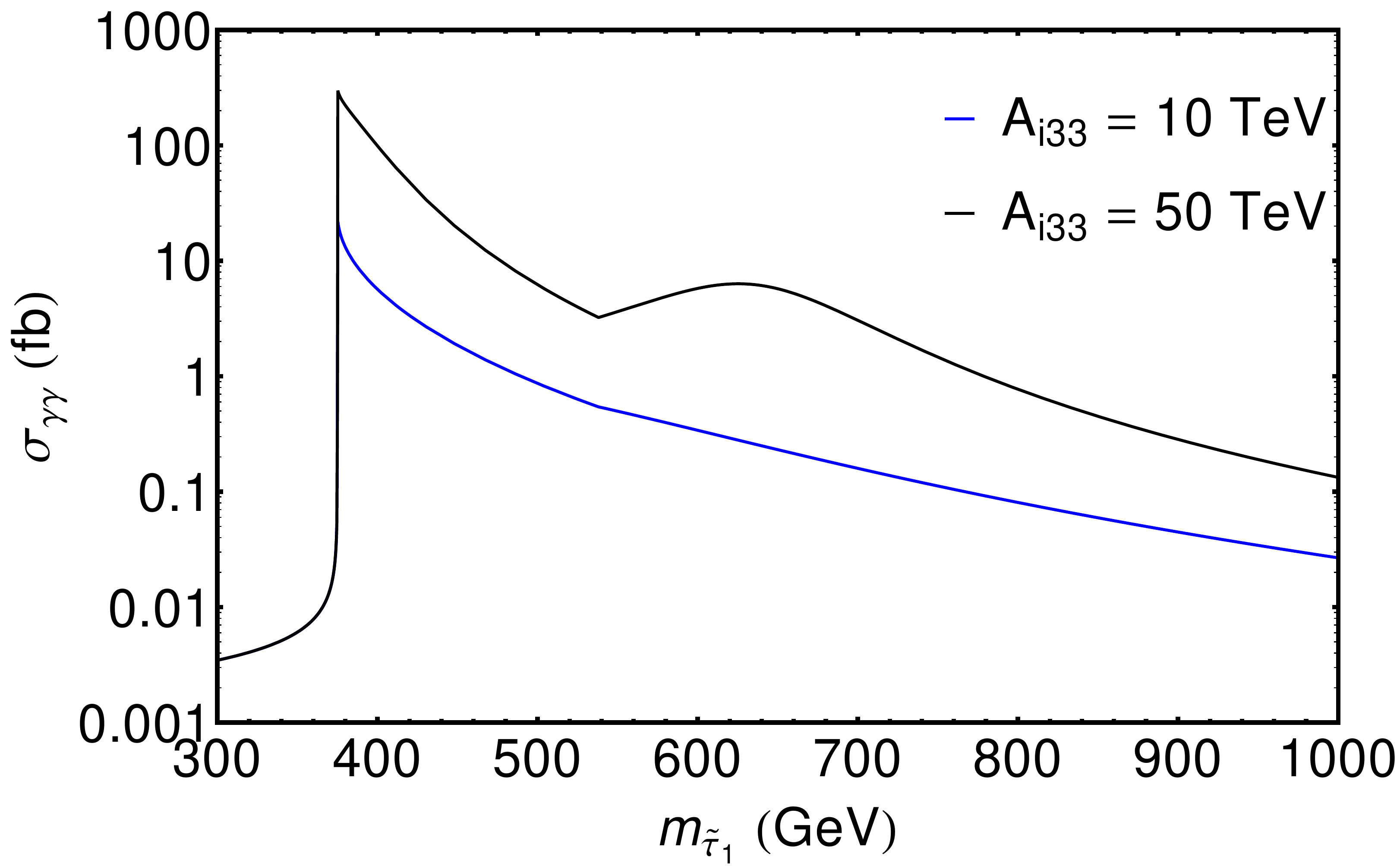}
\caption{Diphoton signal cross section as a function of stau mass, $m_{\tau_{1}}$. The upper panel is for
different RPV couplings $\lambda_{i11}^{\prime}$ with $A_{i33} = 50$ TeV. The lower panel is for $\lambda_{i11}^{\prime} = 0.07$.}
 \label{fig:sigvsmstau}
\end{figure}

A TeV mass sneutrino can be produced at the LHC by the $\lambda^{\prime}_{ijk}L_{i}Q_{j}D_{k}$ term of the superpotential (\ref{SP2}). 
A non-zero RPV coupling  $\lambda_{i11}^{\prime}$ may also give rise to the dijet signals which is constrained by the
LHC observations and is $|\lambda^{'}_{i11}|\leq 0.08$
~\cite{Allanach:2015ixl}. The trilinear vertex $A_{ijk}$ may lead to the decay of sneutrino to charged sleptons. But if the charged sleptons are heavier than the sneutrinos then the sneutrinos will decay to diphotons generated by the scalar loops. The most promising signal at the LHC for observing the TeV scale sneutrinos with large trilinear couplings to scalars can be via TeV scale diphoton resonance. In this section we will show that a diphoton signal with an appreciable cross section which can be seen at LHC can arise from the $R$-parity violating inflation model and this inflation model therefore has a chance of being tested at the LHC. 
Recently it has been pointed out ~\cite{Ding:2015rxx, Allanach:2015ixl} that the sneutrino $\nu_{i}$ can be the origin of a slight excess observed\footnote{With more $\sqrt{s} = 13$ TeV data the ATLAS~\cite{Lenzi:2207286} and CMS~\cite{CMS-PAS-EXO-16-027} collaborations show that the signal no longer persists.} at the LHC in the diphoton invariant mass~\cite{Khachatryan:2016hje, Aaboud:2016tru} if $R$-parity is violated~\cite{Ding:2015rxx, Allanach:2015ixl}. In this scenario the $\sim$ TeV sneutrino can be produced via quark-antiquark interaction through the $R$-parity violating interaction ($\lambda^{\prime}_{ijk}$) mentioned in Eq.~\ref{SP2}. The subsequent decay of the sneutrino to two photons is induced via the stau loop through the RPV soft SUSY breaking term $A_{ijk}$. The relevant Feynman diagrams are shown in Fig.~\ref{fig:feyndiag}. 
Apart from the loop-induced decay to diphoton, sneutrino will also have loop-level decays to $Z\gamma$, $ZZ$, $W^{+}W^{-}$, and tree-level decays to $q\bar{q}$ (we will use $d\bar{d}$ which involves the coupling $\lambda^{'}_{i11}$) and $\tilde{\tau}_{1}\tilde{\tau}_{1}$ if $2m_{\tilde{\tau}_{1}}\leq m_{\tilde{\nu}_{i}}$. In Fig.~\ref{fig:sigvsmstau} we show the cross section for the diphoton signal as a function of stau mass with various $R$-parity violating couplings. 
The SUSY breaking mechanism we used predicts all the scalar masses to be equal at the GUT scale. After RG running down to the electroweak scale the splitting between the stau and sneutrino masses can be in the range which can give the required diphoton cross section, as is presented in Fig.~\ref{fig:sigvsmstau}. For example, a common scalar mass $m_{0} = 600$ GeV at the GUT scale will imply a sneutrino mass $\sim$ 750 GeV and the stau mass $\sim$ 700 GeV at SUSY breaking scale, which can give a diphoton signal cross section $\sigma (pp \rightarrow \phi \rightarrow \gamma \gamma) \sim 10 ~{\rm fb}$, as can be seen from the figure.

\section{Conclusions}
\label{sec:concl}
Plateau inflation in MSSM  with a $D$-flat combination of Higgs fields has been studied in \cite{Chakravarty:2016fin}. In this paper we show that plateau inflation can be achieved in the $R$-parity violating MSSM where the TeV scale sneutrino and charged slepton masses can give testable LHC prediction like a diphoton resonance with significant cross section ($\sigma_{pp \rightarrow \phi \rightarrow \gamma \gamma} \sim 10 ~{\rm fb}$). This signal may show up as TeV scale resonance in the future.
Sneutrino inflation models also have applications in leptogenesis by the Affleck-Dine mechanism \cite{Affleck:1984fy} as has been studied earlier \cite{Garcia:2013bha} and these leptogenesis mechanisms can be studied in our specific model of sneutrino plateau inflation.

\section*{Acknowledgements}
We thank the anonymous referee for his constructive suggestion.

\section*{References}
\bibliographystyle{JHEP}
\bibliography{sneutrino_inflnref.bib}

\end{document}